# Nanodiamonds with photostable, sub-gigahertz linewidths quantum emitters

*Toan Trong Tran[1], Mehran Kianinia[1], Kerem Bray[1], Sejeong Kim[1], Zai-Quan Xu[1], Angus Gentle[1], Bernd Sontheimer[2], Carlo Bradac[1], and Igor Aharonovich[1]*

[1] School of Mathematical and Physical Sciences, University of Technology Sydney, Ultimo, NSW, 2007, Australia
[2] Institute for Physics, Humboldt-University of Berlin, Berlin, Germany

ABSTRACT

Single photon emitters with narrow linewidths are highly sought after for applications in quantum information processing and quantum communications. In this letter, we report on a bright, highly polarized near infrared single photon emitter embedded in diamond nanocrystals with a narrow, sub GHz optical linewidths at 10K. The observed zero phonon line at ~ 780 nm is optically stable under low power resonant excitation and blue shifts as the excitation power increases. Our results highlight the prospect for using new near infrared color centers in nanodiamonds for quantum applications.

TEXT

Solid-state single-photon emitters (SPEs) such as quantum dots or defects in solids are becoming prominent candidates for realization of scalable quantum information technologies[1-4]. In particular, color centers in diamond are one of the most attractive candidates for quantum applications due to their optical stability and availability of spin photon interfaces[5-7].

Among more than 100 luminescent defects in diamond, majority of studies were focused on the negatively charged nitrogen-vacancy (NV). However, the NV center has a major drawback, as only ~ 4% of the photons are emitted into the zero-phonon line (ZPL)[8], which requires sophisticated cavity engineering[9-11] and imposes challenges on its usage in scalable nanophotonics devices. Therefore, other types of defects in diamond with narrow lines and high Debye Waller factors have been recently investigated. These include the silicon vacancy (SiV)[12-16], the germanium vacancy (GeV)[17-19] and other emitters at the near infrared (NIR) spectral range[20-22].

In this work, we explore the optical properties and the linewidth of SPEs embedded in a chemically vapor deposition (CVD) grown nanodiamonds. At room temperature, this emitter have a linewidth of ~ 4 nm and a ZPL at ~ 780 nm, while at cryogenic temperatures, the linewidth reduces ~ 660 MHz, only slightly broader than the Fourier Transform (FT) limited linewidth. These properties of our emitter make them extremely promising for a variety of quantum photonic applications. Furthermore, the sub GHz linewidths from defects in nanodiamonds, rather than in a bulk diamond crystal, offers interesting possibilities to realize hybrid quantum photonic networks[23, 24].

The diamond nanocrystals were grown from detonation nanodiamond seeds (diameter 4-6 nm) that were spin coated on sapphire substrates. The use of sapphire substrates assists in minimizing incorporation of silicon atoms that would otherwise result in undesired silicon doping. The samples were then loaded into a microwave plasma chemical vapor deposition (MPCVD) system and the crystals were grown in mixed gases (hydrogen:methane = 100:1) with a microwave power of 900W, at 60 Torr of atmospheric pressure for 30 minutes. Under these conditions, nanodiamonds with diameters of ~0.3 – 1 µm were grown.

The optical properties of the emitters were studied using a home-built laser scanning confocal photoluminescence (PL) microscope equipped with a continuous wavelength tunable Titanium Sapphire laser (linewidths, 100 KHz) (figure 1a). The sapphire substrate with the grown diamond nanocrystals were mounted onto a cryostat equipped with a high-precision XYZ piezo scanning stage and were cooled to 10K using liquid helium. The excitation and collection were done via a high numerical aperture (NA = 0.9) objective mounted inside the cryostat, creating an excitation and collection spot size of ~ 430 nm. The signal collected from the emitters was analyzed with both a spectrometer equipped with a high-resolution silicon-based charge coupled device (CCD) camera and a Hanbury-Brown and Twiss (HBT) interferometer.

On average, two to three emitters were found in a 60 x 60 µm$^2$ scan area. Bright spots that correlate to single emitters embedded in the nanodiamonds, are shown in figure 1b. Top left inset of fig 1b, is the scanning electron microscope (SEM) image of a typical nanodiamond grown under our experimental conditions. Cross-sectional analysis of each spot shows a Gaussian profile with a full-width-at-half-maximum (FWHM) of 0.44 µm, very close to the diffraction-limited point spread function of our confocal system (bottom right inset). Detailed spectroscopic study of the sample revealed that the nanodiamonds host narrowband color centers with ZPL in the range of 756–786 nm (supporting information figure S1). For the remaining of the manuscript, we focus on a particular line at 780 nm. The line was selected arbitrary, with a goal to study emitters that emit further in the NIR. Consequently, figure 1c shows the narrow ZPL at 780 nm with a FWHM of ~ 0.16 nm that corresponds to our spectrometer resolution. Inset is the spectrum of the same emitter recorded over broader spectral range. The spectra were recorded at 10 K. From the spectrum, we deduced a Debye-Waller (DW) factor, $DW = I_{ZPL}/I_{tot} = 0.87$, a higher value than that of nitrogen-vacancy color center in diamonds (0.04) and comparable with the SiV and GeV defects).

To verify that the emitter is indeed a single photon emitter, a second-order autocorrelation function, $g^2(\tau)$, was recorded and is shown in figure 1d. The data is fitted by employing a standard three-level model for the color center: $g^{(2)}(\tau) = 1 - (1+a)e^{-\tau/t_1} + ae^{-\tau/t_2}$ where $a$ is the bunching factor, while $t_1$ and $t_2$ are the lifetimes of the excited and metastable state, respectively. The fit yields a value of 0.26 for $g^{(2)}(0)$, indicating the quantum nature of the emission. The saturation curves of the SPE and its polarization properties are shown in the supporting information (figure s2).

To gain more information about the coherent properties of the SPEs, and unveil its natural linewidth, resonant exciton measurements were performed. While cross polarization schemes are often used for studying quantum dots[25], they were not practical in our measurement due to the high scattering from the diamond nanocrystals. To filter the exciting laser, we used a long-pass filter to collect only the emitter's phonon side band (PSB). Figure 2a shows a resonant excitation spectrum recorded at 5 μW with 5 GHz scan range and 70 MHz resolution of the same SPE as shown in figure 1 ( center wavelength at 779.61 nm or 384.54 THz). A single peak was clearly observed. The data is fit with a Lorentzian profile, producing a full-width-at-half-maximum (FWHM) of 660 MHz and as a comparison with a Gaussian line shape, that results in a linewidth of 800 MHz. The Lorentzian line shape results in a better fit when compared to the Gaussian fit, suggesting that the optical linewidth is less prone to spectral diffusion[21, 26, 27], as will be discussed later.

The excited state lifetime of the emitter was measured using a pulsed laser to find out whether the emitter's linewidth is Fourier Transformed (FT) limited. The results are shown in Figure 2b. Using a single exponential fit, the lifetime of the excited state of the emitter was determined to be $\tau_1 = 1.1$ ns. By applying the expression: $\gamma = 1/2\pi\tau_1$ where $\gamma$ and $\tau_1$ are the FT limited linewidth (natural linewidth) and the excited state's lifetime of the emitter, respectively, we estimated a value for $\gamma$ of 145 MHz. This means that the emitter's linewidth is ~ 4.5 times broader than the expected value.

The linewidth broadening can arise from numerous factors including ultrafast spectral diffusion due to interaction of the strong emitter dipole with fluctuating electric field from surrounding defects (inhomogeneous broadening), or alternatively, a homogeneous broadening due to phonon coupling. In addition, spectral diffusion often results in intensity fluctuations, associated with slow frequency jumps, which were not observed in our case. Combined with a more favorable Lorentzian fit, we therefore conclude that the line broadening is predominantly due to phonon interactions. Similar behavior was also reported for the SiV[15, 28] and the GeV[17] in diamond. Despite the line broadening, achieving a sub GHz linewidths from single emitters in nanodiamonds, particularly at the NIR spectral range is valuable, as it opens excellent pathways to couple these emitters to high quality optical resonators and photonic cavities[23, 24].

Next, we measured the emitter's linewidth off resonantly as a function of laser power to determine the photostability of the emitter under increasing excitation power. Figure 3 (a, b) show the optical stability of the emitters under a low 300 µW and high 3 mW excitation laser power, respectively. For these measurements, a total of 200 PL spectra were acquired at intervals of 200 ms. At low power (300 µW), almost no spectral diffusion or blinking to a different frequency was witnessed (within the spectrometer resolution). On the other hand, at high excitation power of ~ 3 mW the spectral fluctuation of the ZPL were noticeable. Spectral jumps as large as ~1.5 nm away from the ZPL position were observed, several orders of magnitude higher than the measured FWHM of ~ 660 MHz. This suggests that the emitter is highly susceptible to the strength of the laser electromagnetic field, and it may possess a linear permanent dipole behavior that may result in spectral diffusion under increased excitation power. Additionally, an increased pumping intensity may result in a photo-ionization, similarly to what occurs with the NV center[29]. Photoionization will result in a frequency drift, and can be evidenced as blinking. Note, however, that the spectral jumps occurred on second time scale, which means techniques such as dynamic stabilization using applied electric fields can be employed to stabilized the emitter under high excitation power.[30]

In addition to the power induced blinking, the ZPL exhibits broadening as a function of excitation power. Figure 3c shows several spectra from the same SPE under increased excitation power which reveals power broadening and a blue shift in ZPL spectral positions. Figure 3d presents a plot of the FWHM values as a function of power, showing a good fit with a logarithmic function. These measurements are in agreement with previous optical studies on carbon nanotubes, suggesting that the broadening arises from an increase of the local temperature induced by an increase in power of the excitation laser. [31, 32]

Finally, we discuss the potential origin of the emitters. Amongst color centers in diamond, only SiV, GeV and an unknown NIR defect[21] exhibited narrow lines amenable to resonant excitation. Our emitters cannot be attributed to the SiV, since we did not observe splitting into 4 spectral lines at cryogenic temperatures – a typical signature of a single SiV defect. The emitter's ZPL at ~ 780 nm is also far from the standard SiV emission centered at 738 nm.[14] The defects can be attributed to Cr related defects, as the sample was grown on sapphire, in a similar procedure as was reported in previous works[33], although previous measurements from these emitters revealed a broader GHz lines[21]. Another viable explanation is the recently discovered narrowband emitters that appear at secondary nucleation cites and extended defects in CVD grown nanodiamonds.[34] As seen from SEM image in figure S1a, some nanodiamonds possess such defects, and therefore give rise to the narrowband PL lines, as indeed observed in our experiments. While we cannot conclusively identify the origin of the emitters, or its crystallographic structure, the scalable production of these emitters from a standard growth procedure enables unique opportunities to explore quantum optics experiments and future applications with single emitters in nanodiamonds.

In conclusion, our study shows promising optical properties of SPEs with ZPLs at the NIR (~ 780 nm). The resonant excitation measurements suggest the SPEs have FWHM of ~ 660 MHz, with a homogenious broadening due to interactions with phonons. The studied emitter does not show any blinking under resonant and low power excitation. The SPEs exhibits high DW factors exceeding 0.85, with fast florescence lifetime and fully polarized emission. Our results should stimulate more studies into the promising attributes of NIR emitters in nanodiamonds and their use in quantum photonics applications.

**Acknowledgments**

Financial support from the Australian Research Council (DE130100592), FEI Company, the Asian Office of Aerospace Research and Development grant FA2386-15-1-4044 are gratefully acknowledged. This research is supported in part by an Australian Government Research Training Program (RTP) Scholarship.REFERENCES

1. P. Lodahl, S. Mahmoodian, and S. Stobbe, "Interfacing single photons and single quantum dots with photonic nanostructures," Reviews of Modern Physics **87**, 347-400 (2015).
2. I. Aharonovich, D. Englund, and M. Toth, "Solid-state single-photon emitters," Nat. Photonics **10**, 631-641 (2016).
3. H. Wang, Y. He, Y.-H. Li, Z.-E. Su, B. Li, H.-L. Huang, X. Ding, M.-C. Chen, C. Liu, J. Qin, J.-P. Li, Y.-M. He, C. Schneider, M. Kamp, C.-Z. Peng, S. Höfling, C.-Y. Lu, and J.-W. Pan, "High-efficiency multiphoton boson sampling," Nat Photon **advance online publication**(2017).
4. M. J. Holmes, K. Choi, S. Kako, M. Arita, and Y. Arakawa, "Room-Temperature Triggered Single Photon Emission from a III-Nitride Site-Controlled Nanowire Quantum Dot," Nano Lett. **14**, 982-986 (2014).
5. E. Togan, Y. Chu, A. S. Trifonov, L. Jiang, J. Maze, L. Childress, M. V. G. Dutt, A. S. Sorensen, P. R. Hemmer, A. S. Zibrov, and M. D. Lukin, "Quantum entanglement between an optical photon and a solid-state spin qubit," Nature **466**, 730-U734 (2010).
6. A. Sipahigil, R. E. Evans, D. D. Sukachev, M. J. Burek, J. Borregaard, M. K. Bhaskar, C. T. Nguyen, J. L. Pacheco, H. A. Atikian, C. Meuwly, R. M. Camacho, F. Jelezko, E. Bielejec, H. Park, M. Lončar, and M. D. Lukin, "An integrated diamond nanophotonics platform for quantum optical networks," Science (2016).
7. H. Bernien, B. Hensen, W. Pfaff, G. Koolstra, M. S. Blok, L. Robledo, T. H. Taminiau, M. Markham, D. J. Twitchen, L. Childress, and R. Hanson, "Heralded entanglement between solid-state qubits separated by three metres," Nature **497**, 86-90 (2013).
8. F. Jelezko and J. Wrachtrup, "Single defect centres in diamond: A review," Physica Status Solidi a-Applications and Materials Science **203**, 3207-3225 (2006).
9. B. J. M. Hausmann, B. Shields, Q. M. Quan, P. Maletinsky, M. McCutcheon, J. T. Choy, T. M. Babinec, A. Kubanek, A. Yacoby, M. D. Lukin, and M. Loncar, "Integrated Diamond Networks for Quantum Nanophotonics," Nano Lett. **12**, 1578-1582 (2012).
10. S. L. Mouradian and D. Englund, "A tunable waveguide-coupled cavity design for scalable interfaces to solid-state quantum emitters," APL Photonics **2**, 046103 (2017).
11. D. Englund, B. Shields, K. Rivoire, F. Hatami, J. Vuckovic, H. Park, and M. D. Lukin, "Deterministic Coupling of a Single Nitrogen Vacancy Center to a Photonic Crystal Cavity," Nano Lett. **10**, 3922-3926 (2010).
12. B. Pingault, D.-D. Jarausch, C. Hepp, L. Klintberg, J. N. Becker, M. Markham, C. Becher, and M. Atatüre, "Coherent control of the silicon-vacancy spin in diamond," arXiv preprint arXiv:1701.06848 (2017).
13. T. Müller, C. Hepp, B. Pingault, E. Neu, S. Gsell, M. Schreck, H. Sternschulte, D. Steinmüller-Nethl, C. Becher, and M. Atatüre, "Optical signatures of silicon-vacancy spins in diamond," Nat Commun **5**(2014).
14. N. Elke, S. David, R.-M. Janine, G. Stefan, F. Martin, S. Matthias, and B. Christoph, "Single photon emission from silicon-vacancy colour centres in chemical vapour deposition nano-diamonds on iridium," New J. Phys. **13**, 025012 (2011).6

**Figures**

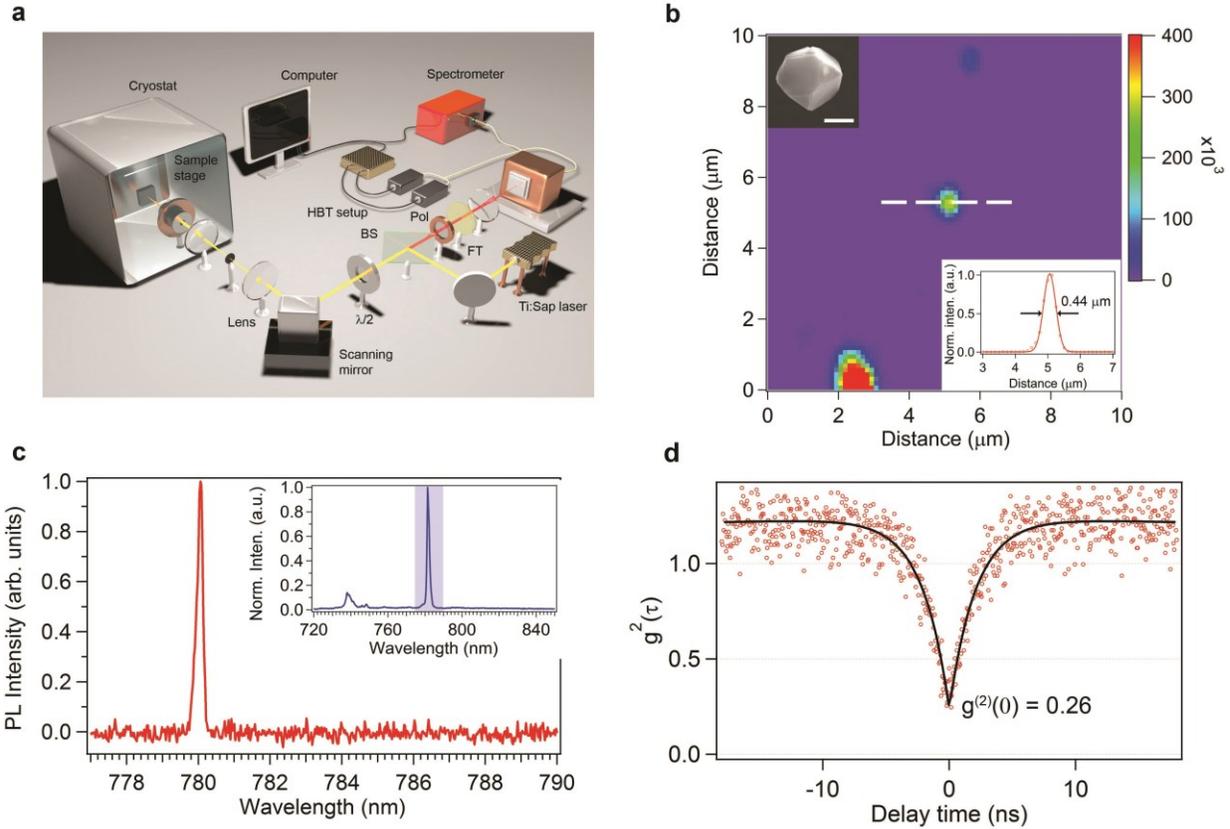

**Figure 1. a)** Cryogenic confocal PL setup. HBT: Hanbury-Brown and Twiss; BS: beamsplitter; FT: band-pass or long-pass filters; λ/2: half waveplate; Pol: linear polarizer. **b)** Confocal PL map recorded with 700-nm laser excitation at 300 μW. The bright spot corresponds to a nanodiamond hosting a single emitter. The bottom inset shows the cross-sectional intensity analysis (dotted line) revealing a FWHM of 0.44 μm, consistent with emission from a point-source. The top inset shows an SEM image of a representative grown nanodiamond (scale bar 500 nm). **c)** Normalized PL spectrum taken from the same color center (red trace) with a 1800g/mm grating. The inset shows a larger spectrum window taken from the same emitter (with a 300g/mm grating). The semitransparent blue window indicates the band-pass filter transmission range used in the measurements of the photon second order autocorrelation function, fluorescence saturation, and lifetime to minimize the background PL contribution. **d)** Second-order autocorrelation function (red open circles) acquired for the emitter with a band pass filter (785 ± 22) nm (blue window in (c)). The black solid line is the fitting (see main text) for the $g^{(2)}(0)$ function. The value of 0.26, without any background correction, indicates that the emission is from a single emitter. The measurements in (b) and (c) were carried out at 10 K, while the measurements in (d) at 80K.

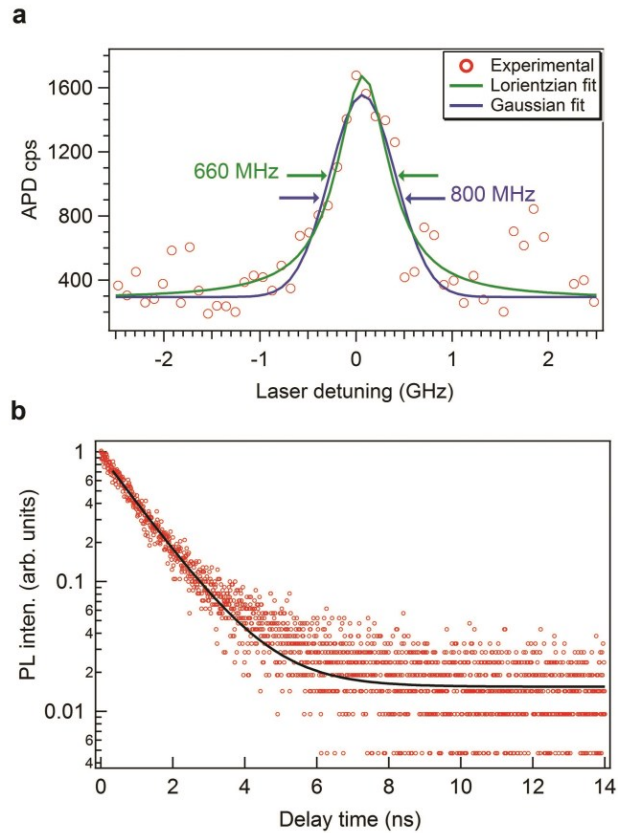

**Figure 2. a)** Resonant photoluminescence excitation measurements on the single emitter with a ZPL peak of 779.61 nm. The excitation power used was 5 µW. Only photons from the PSB were collected using a long pass filter. The experimental data is plotted as open red circles. The data was fit with either a Lorentzian (green line) or Gaussian (blue line) function. The measurement was done at 10 K. **b)** Time-resolved PL measurements (red open circles) of the same single emitter measured at room temperature. A single-exponential fit gives rise to a lifetime of 1.1 ns for the emitter's excited state. The measurement was done with a 675-nm pulsed laser (100 µW, 10 MHz repetition rate, 100-ps pulse width).

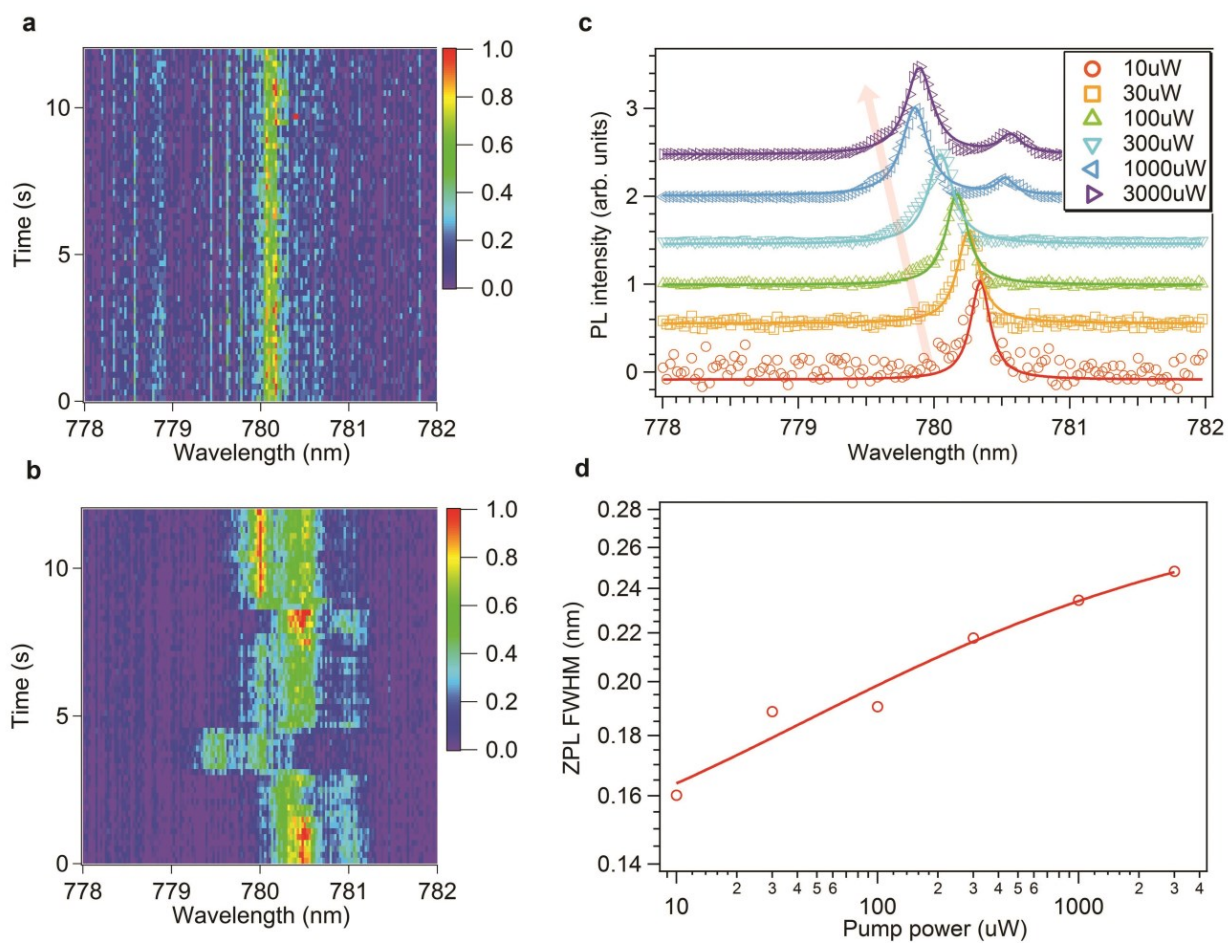

**Figure 3.** Spectral stability measurements taken from the emitter at 300 μW **(a)** and 3mW **(b)**. All the spectra shown were normalized. **c)** Power-induced linewidth broadening measurements for the same emitter with laser power increasing from 10 μW to 3mW. The open markers and solid lines are experimental and fit data, respectively. The red semitransparent arrow serves as a guide-to-the-eye for the shift in the spectra. **d)** Lorentzian fitted FWHM of the linewidths of the emitter from (c). The spectral broadening is evident. The data is fit with a log function.